\documentclass[english]{article}
\usepackage[T1]{fontenc}
\usepackage[latin9]{inputenc}
\usepackage{babel}
\usepackage{graphicx}
\usepackage{caption}
\usepackage{subcaption}
\begin{document}

\title{Roads and cities of $18^{th}$ century France}

\author{Julien Perret\textsuperscript{1{*}}, Maurizio Gribaudi\textsuperscript{2}, Marc Barthelemy\textsuperscript{3,4}}

\maketitle
1. COGIT, IGN. 73 avenue de Paris, 94165 Saint-Mande Cedex, France. 2. LaD\'eHiS, EHESS. 190-198 avenue de France, 75013 Paris, France. 
3. IPhT, CEA. Orme-des-Merisiers, 91191 Gif-sur-Yvette, France. 4. CAMS, EHESS. 190-198 Avenue de France, 75013 Paris, France.\\
{*} Corresponding author (julien.perret@gmail.com)
\begin{abstract}
The evolution of infrastructure networks such as roads and streets are of utmost importance to understand the evolution of urban systems.
However, datasets describing these spatial objects are rare and sparse.
The database presented here represents the road network at the french national level described in the historical map of Cassini in the $18^{th}$ century.
The digitization of this historical map is based on a collaborative methodology that we describe in detail.
This dataset can be used for a variety of interdisciplinary studies, covering multiple spatial resolutions and ranging from history, geography, urban economics to network science.
\end{abstract}

\section*{Background \& Summary}
\begin{figure}[htb]
  \centering
  \includegraphics[width=\textwidth]{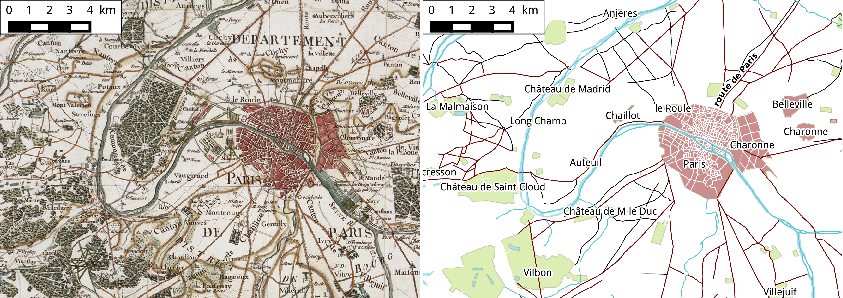}
  \caption{Part of the Cassini map of Paris and its digitization. The map is produced by EHESS, CNRS and BnF~\cite{cassini} and can be freely accessed by web service~\cite{geoportail}.}
  \label{fig:paris}
\end{figure}

Triggered by recent, powerful digitization techniques, there is a huge interest in historical data, in particular when they allow to track temporal changes at different spatial scales.
Such projects comprise for example the NYPL initiative~\cite{NYPL}, the digitization of the road network of a region in Italy~\cite{Strano2012}, of Paris over 200 years~\cite{Barthelemy2013}, and the digitization of ancient French forests~\cite{Dupouey2007,Vallauri2012}.
New historical datasets extracted from maps allow researchers to study the time evolution of urban systems, to extract stylized facts, and for the first time to test theoretical ideas and models.
Historical datasets of road networks allow to study territorial evolutions at different scales and to build tools to accurately answer theoretical questions.
In particular, one can ask about the impact of the road network on subsequent urbanization, the correlation between the location of an entity (such as a city, town, etc.) and socio-economical indicators such as population or importance in the trade network, immigration, etc.
More generally, such historical datasets are of interest to a wide variety of scientists comprising historians, geographers, mathematicians, archeologists, geo-historians, geomaticians, and computer scientists~\cite{Masucci2013,Wang2015,Gribaudi2014,Porta2014}.
The digitization of historical sources is usually done locally by researchers for their immediate research needs without sharing their work and results with others.
In contrast, we believe that it is essential to build a platform to share our work, but also to have a collective control over the production process of the data, its transformation and its analysis.

Operations such as scanning, georeferencing and digitization of historical sources imply several and delicate choices that should be documentated and tracked.
Historical sources might have deformations originating from aging.
Their georeferencing carries its own deformations which have to be minimized in order for the sources to remain legible.
Our approach consists in taking these geometric displacements into account after the digitization process using spatial data matching tools~\cite{Walter1999} to find corresponding entities in consecutive data sets.
Such tools should allow researchers to control and take into account the imperfections of the data throughout their analysis~\cite{Olteanu2008}.
This way, we can reduce the impact of the georeferencing in the matching process and the analysis.
Furthermore, opendata and open source tools provide the scientific community with the ability to control, track and reproduce the results at every stage.

With these ideas in mind, we developed a collaborative way to digitize the Cassini map of the 18th century (see Figure~\ref{fig:paris} for a visualization of a small subset of the map and the corresponding digitized data).
This map is the first one that restitutes with geometrical precision the entire French territory in the second half of the eighteenth century at a scale of 1/86 000.
First conceived in the late $17^{th}$ century, this work was made possible by the development of geodesic triangulation techniques and their generalization.
The determination of the Paris meridian and the establishment of a single framework for all triangulations of France (1744) provided the reference needed for putting together several local maps~\cite{Maraldi1744}.
In 1747 C\'esar-Fran\c{c}ois Cassini de Thury was formally commissioned by Louis XV to draw the entire map of France showing the entire kingdom but also finer details.
Cassini and his engineers divided the French territory in a grid of 180 rectangles with a size of about 80 km $\times$ 50 km which lead to as many maps printed on sheets of size 104 cm $\times$ 73 cm.
Due to financial difficulties, the Revolution and regime changes, the constitution of this map was delayed and it is not before 1815 that the last sheets were released, under the direction of Jean-Dominique Cassini, son of C\'esar-Fran\c{c}ois.

The maps that serve as a basis for our work is the digital copy of the so-called ``Marie-Antoinette'' version, commissioned in 1780 by the queen.
These maps were completed, corrected and updated in the subsequent years.
For example, the map of the Paris region which was initially drawn between 1749 and 1755, and published the first time in 1756, displays clear signs of corrections made during the post-revolution period with the introduction of administrative divisions created during the Republic in 1790.

An important part of the project was therefore to analyze each sheet, to give a precise date of its drawing and to provide an assessment of its accuracy.
This was done by comparing different printed and dated versions, and many minutes and notes from the National Institute of Geographic and Forest Information (IGN) archives.
The main work was however (see Methods) to analyze and vectorize a large number of features of the Cassini map such as roads, water networks, towns and villages, forest and crops, industrial and administrative structures.
The digitized data have been made available on a dedicated geo-historical portail~\cite{geohistoricaldatawebsite}.
These different features put together under a digital form give us a detailed picture of the french territory in the second half of the eighteenth century. 

\section*{Methods}
The digitization of the Cassini maps and, in particular, of its road network, was achieved in a collaborative way using a shared PostgreSQL~\cite{postgresql} database and its spatial extension PostGIS~\cite{postgis}.
GIS editing tools such as Quantum GIS~\cite{qgis} were used to remotely digitize the objects using a WMTS (Web Tile Map Service) layer provided by IGN~\cite{cassini} as background. Details on the methods used to produce the georeferenced map are available on a dedicated website~\cite{cassiniwebsite}.
This way, several operators have been able to digitize data simultaneously on the same database.
In order to provide consistent data records, data specifications were proposed as a result of an important collaborative work.
Nevertheless, as the specifications were enhanced during the digitization process, local variations in the capture of several attributes might be found (the attribute ``bordered'' was added after a few months of digitization for instance).
Further work will focus on the consistency of the data (both for attributes and geometries).

\begin{figure}[htb]
  \centering
  \includegraphics[width=\textwidth]{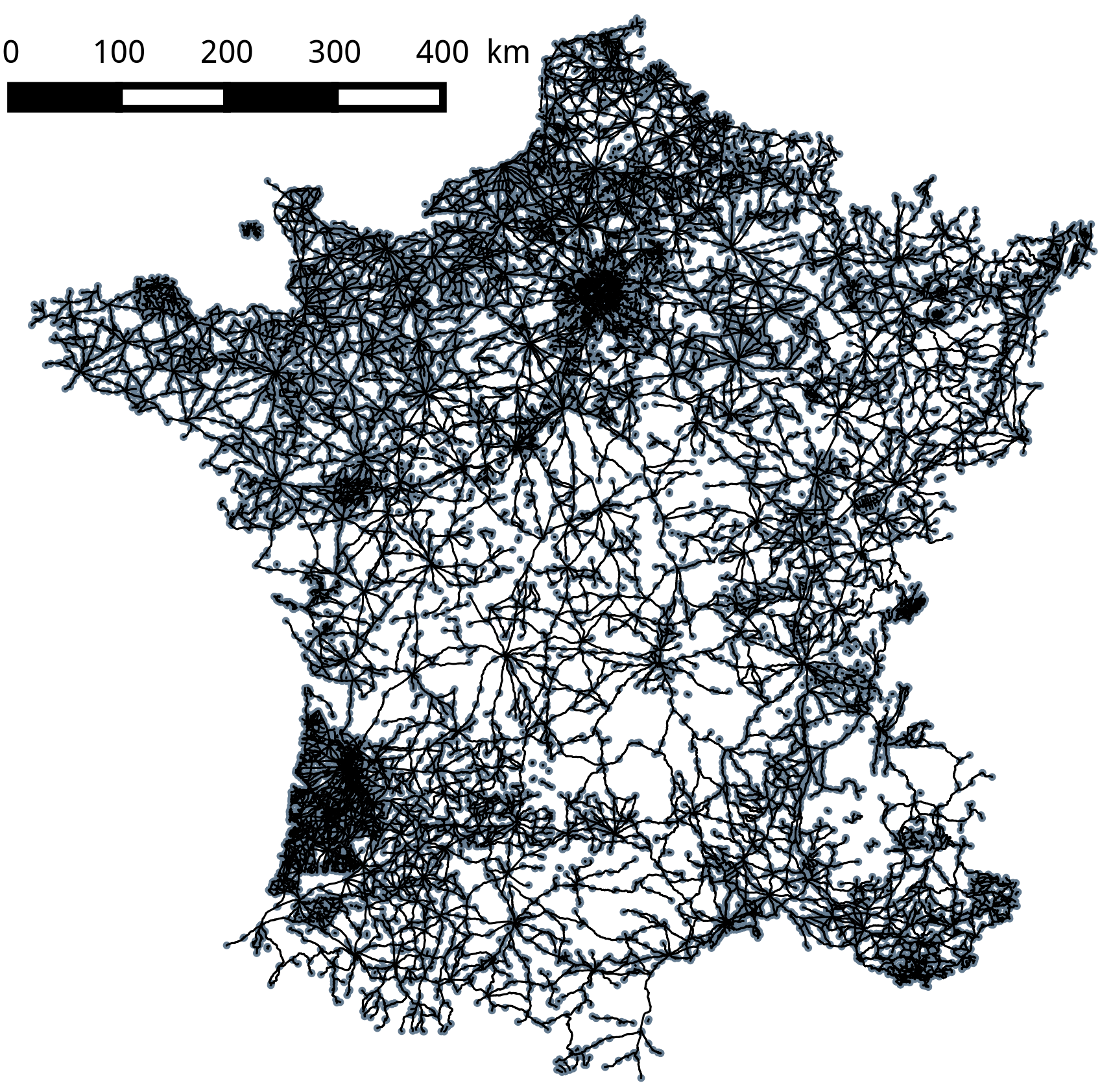}
  \caption{The digitized $18^{th}$ century french road network.}
  \label{fig:french_network}
\end{figure}

\begin{figure}[htb]
  \centering
  \includegraphics[width=\textwidth]{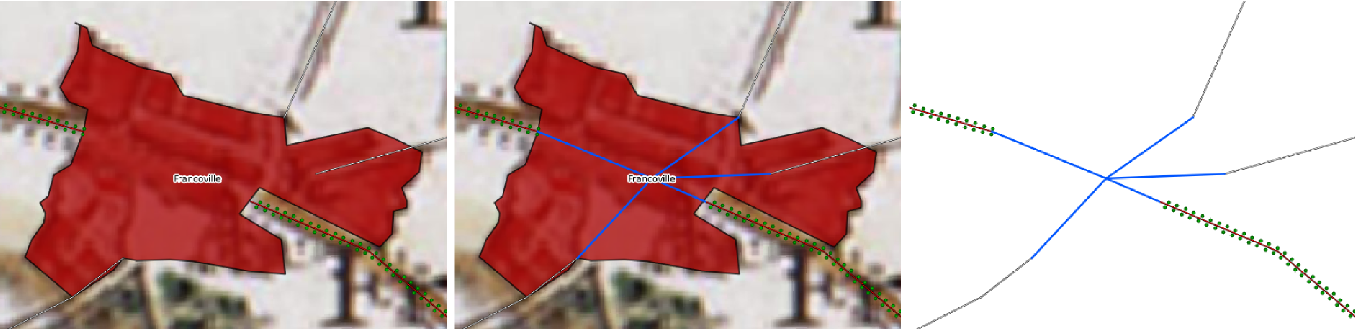}
  \caption{Construction of ``fictive'' edges in cities: the digitized edges of the road network connected to the city are linked by the created edges (in blue).}
  \label{fig:city_fictive_edge}
\end{figure}

An important aspect of the Cassini dataset is the fact that the Cassini map was not homogeneously drawn 
(different sheets might show different levels of detail as seen in Figure~\ref{fig:french_network}) or conceived as a road map~\cite{Pelletier2002}.
Hence, one has to be careful when studying the road network extracted from it~\cite{Bonin2014}.
Specifically, the road network inside most cities was not drawn in the map.
An automatic process is therefore proposed to create so-called ``fictive'' edges inside cities allowing to link all roads leading them.
As shown in Figure~\ref{fig:city_fictive_edge}, a node representing the city is created at its centroid (or rather at the centroid of the geometry representing its boundary in the map) and edges are created to connect this node to the edges ending in the city.
Furthermore, in order to speed up the digitizing process, some roads have been captured as continuous strokes rather than by topological road segments: some users digitized entire roads instead of stopping the capture at each road intersection.
We therefore use the PostGIS topology engine~\cite{postgistopology} to convert the digitized strokes into a topological network.
This process uses a distance threshold to merge points closer than the given threshold and thus allows for the correction of minor shifts between points and a second threshold for to collect all nodes in the neighboorhood of a city.
The thresholds used in the current export are 10 meters and 20 meters respectively.
The digitized roads and cities are also provided in the export and the code for the topological export is available~\cite{cassinitopology}.

\section*{Data Records}
The data records contain the roads and cities as captured (the names of the attributes have been translated though) and the topological nodes, edges and faces.
We propose five shapefiles (which each actually refer to four files with .shp, .dbf, .shx and .prj extentions) and two CSV files containing simplified versions of the nodes and edges. 
The dataset is stored at the Harvard Dataverse (Data Citation 1).

\subsection*{Roads (france\_cassini\_roads.shp)}
This file contains the roads represented in the Cassini maps.
It includes the following attributes: 
\begin{itemize}
\item{\emph{id:}} the (unique) identifier for each road segment (integer);
\item{\emph{geometry:}} the geometry of the segment (linestring) in RGF93 / Lambert-93 (EPSG:2154).
\item{\emph{type:}} the type  of  road  or  connexion  as  represented  in  the  map:  either  ``red'', ``white'', ``trail'', ``forest'', ``bridge'', ``ferry'' or ``gap''. These  values  refer  respectively  to  main  roads,  secondary roads, trails, forest trails, bridges, tubs, and shifts between sheets (string).
\item{\emph{name:}} the name of the segment when it has one (string). 
\item{\emph{uncertain:}} whether the nature of the segment is difficult to clearly  identify in the map (boolean).
\item{\emph{bordered:}} whether the segment is bordered by trees (boolean).
\item{\emph{comments:}} comments left by our contributors when the object raises specific questions (string). 
\end{itemize}

\subsection*{Cities (france\_cassini\_cities.shp)}
This file describes some of the main types of land use identifiable in the maps. 
\begin{itemize}
\item{\emph{id:}} the (unique) identifier for each object (integer). 
\item{\emph{geometry:}} the geometry of the object (multipolygon) in RGF93 / Lambert-93 (EPSG:2154). 
\item{\emph{type:}} the type of object: ``city'', ``town'', ``domain'', ``fort'' (string), respectively  for  cities, towns, domains and forts.
\item{\emph{name:}} the name of the land element when it has one (string).
\item{\emph{fortified:}} is the city fortified? (boolean). Can only be true if the type is ``city''.
\item{\emph{comments:}} comments  left  by   our  contributors  when  the  object  raises  specific  questions (string).
\end{itemize}

\subsection*{Topological Nodes (node.shp)}
\begin{itemize}
\item{\emph{id:}} the (unique) identifier for each object (integer). 
\item{\emph{geom:}} the geometry of the object (point) in RGF93 / Lambert-93 (EPSG:2154). 
\item{\emph{city\_id:}} identifier of the city it lies in (from france\_cassini\_cities.shp) 
\item{\emph{city\_name:}} the name of the city (from france\_cassini\_cities.shp) 
\item{\emph{city\_type:}} the type of the city (from france\_cassini\_cities.shp) 
\item{\emph{component:}} the identifier of the connected component the node belongs to (integer)
\end{itemize}
 
\subsection*{Topological Edges (edge.shp)}
Edges are not oriented so the start and end nodes are arbitrary.
Nevertheless, they are consistent with the order of the points in the geometry of the edge (the start node position is the first point of the geometry of the edge).
When the edge is built from a road, it holds the identifier of this road.
Its type is also given for convenience but is recoverable by join (combining the Edge table with the type from the roads table by using the common identifier road\_id).
Note that ``fictive'' edges do not hold such an identifier.
Furthermore, in cases where multiple roads are merged into the same edge, the identifier is arbitrary.
\begin{itemize}
\item{\emph{id:}} the (unique) identifier for each object (integer). 
\item{\emph{geom:}} the geometry of the object (linestring) in RGF93 / Lambert-93 (EPSG:2154). 
\item{\emph{start\_node:}} identifier of the initial node of the edge (from node.shp) 
\item{\emph{end\_node:}} identifier of the final node of the edge (from node.shp) 
\item{\emph{road\_id:}} identifier of the road it stems from (from france\_cassini\_roads.shp) 
\item{\emph{road\_type:}} type of the road(from france\_cassini\_roads.shp) 
\item{\emph{length:}} length of the edge (meters) 
\item{\emph{component:}} the identifier of the connected component the edge belongs to (integer)
\end{itemize}
 
\subsection*{Topological Faces (face.shp)}
As the resulting network is a planar graph (\emph{i.e.} a graph that can be embedded in the plane), the faces (\emph{i.e.} the regions bounded by edges) are also provided.
\begin{itemize}
\item{\emph{id:}} the (unique) identifier for each object (integer). 
\item{\emph{geom:}} the geometry of the object (polygon) in RGF93 / Lambert-93 (EPSG:2154). 
\end{itemize}
 
\subsection*{Simplified Topological Nodes (node.csv)}
This file contains the same nodes as node.shp but in a different easily accessible format.
The position of the roads is given in lat/long.
\begin{itemize}
\item{\emph{id:}} the (unique) identifier for each object (integer)
\item{\emph{lat:}} the latitude of the node in WGS 84 (EPSG:4326)
\item{\emph{long:}} the longitude of the node in WGS 84 (EPSG:4326)
\item{\emph{city\_id:}} the identifier of the city it lies in (from france\_cassini\_cities.shp) 
\item{\emph{city\_name:}} the name of the city (from france\_cassini\_cities.shp) 
\item{\emph{city\_type:}} the type of the city (from france\_cassini\_cities.shp) 
\end{itemize}
 
\subsection*{Simplified Topological Edges (edge.csv)}
This file contains the same edges as edge.shp without the geometry.
It is therefore a simplified version.
The length of the edge is the cartesian 2D length of the geometry (a linestring, i.e. a sequence of line segments) from edge.shp computed using PostGIS funtion \emph{ST\_Length}.
\begin{itemize}
\item{\emph{id:}} the (unique) identifier for each object (integer)
\item{\emph{start\_node:}} identifier of the initial node of the edge (from node.shp) 
\item{\emph{end\_node:}} identifier of the final node of the edge (from node.shp) 
\item{\emph{road\_id:}} identifier of the road it stems from (from france\_cassini\_roads.shp) 
\item{\emph{road\_type:}} type of the road(from france\_cassini\_roads.shp) 
\item{\emph{length:}} length of the edge (meters)
\end{itemize}

\section*{Technical Validation}
\subsection*{Topological Validation}
The topology created using PostGIS Topology is first validated by the same tool and the provided function \emph{ValidateTopology} without error.
This function checks for several errors including crossing edges, and mismatching edge/node topology.

Furthermore, we compute the number of input edges corresponding to the edges of the final network.
This allows us to identify the duplicated edges, i.e. the edges in the final network which correspond to multiple edges in the input data. These duplicated edges usually correspond to digitization errors and are used to manually validate the digitized data. The latest version (V5) of the topology does not contain any duplicated edge.

\subsection*{Connected Components Validation}
\begin{figure}[!htb]
  \centering
\includegraphics[width=0.8\textwidth]{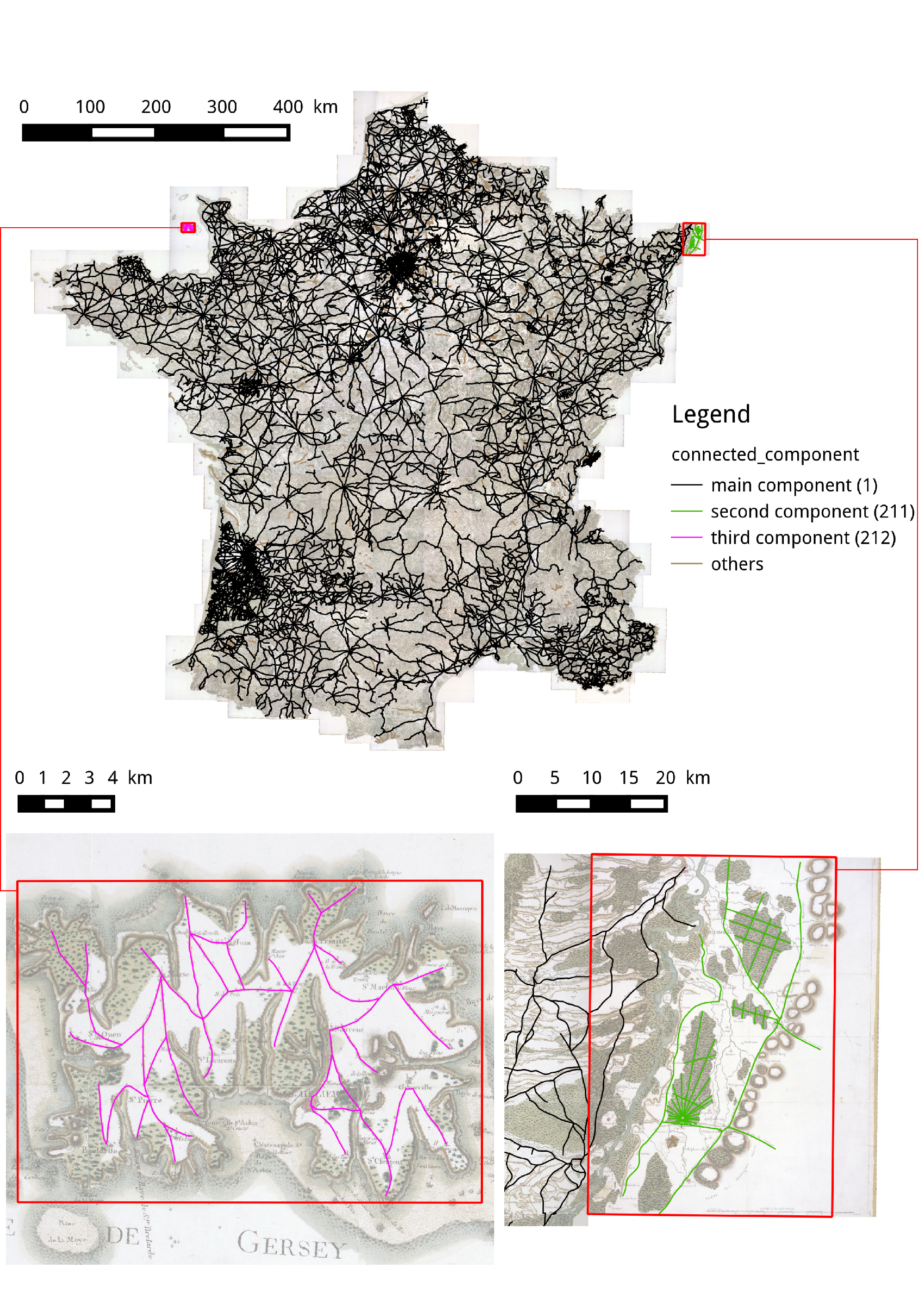}
  \caption{The three largest connected components of the network.}
  \label{fig:components}
\end{figure}

The second validation consists in computing and analysing the connected components of the network.
Indeed, such a road network should essentially be connected and small connected components are unlikely (they would mean small 'islands' disconnected from the rest of the network).
Our network contains 1274 connected components.
The largest component is about 110,000 kilometers in length (more than 96\% of the total length of the network) whereas the smallest is about 100 meters.
Figure~\ref{fig:components} shows the three largest connected components in the network.
Note that the second largest component is at the very edge of the map (in Germany) and is not visually connected to the network in the map.
Finally, the third largest component is the Jersey island.
Other large components represent other islands but also forests which paths are represented (and thus digitized) but rarely connected to the road network.
The smallest components represent isolated features such as bridges.
They can also correspond to digitization errors and the connected components can be used as a tool for data correction.

\subsection*{Collaborative Validation}
Our third validation method is still ongoing work.
It was inspired by the ``Building Inspector''~\cite{buildinginspector}, developped by NYPL and used for the validation of buildings automatically vectorized from insurance maps.
With the help of NYPL, we adapted this tool to collaboratively validate and correct our digitized data.
The resulting application, ``L'Arpenteur Topographe''~\cite{arpenteurtopographe} is being tested on the digitized cities.
The code of the application (from NYPL and our contributions) is available online~\cite{arpenteurtopographe-code}.
Further tests should be carried out on other objects in the future.
Further work will also focus on better handling the interaction between the collaborative digitization process (using desktop or online GIS tools) and the collaborative validation, correction and enrichment processes such as in ``L'Arpenteur Topographe''.


\section*{Acknowledgements}
The digitization of the Cassini maps is the result of the collective work of the following group of colleagues as much as it is the work of the authors (in alphabetical order)~:
N. Abadie (IGN), S. Baciocchi (EHESS), C. Bertelli (Charta s.r.l.), O.  Bonin (IFSTTAR), P. Bordin (Geospective), B. Costes (IGN), P. Cristofoli (EHESS), B. Dumenieu (IGN/EHESS), J. Gravier (Geographie-Cit\'es), J.-P. Hubert (IFSTTAR), P.-A. Le Ny (Le Ny Conseil), E. Mermet (EHESS), C. Motte (EHESS), M. Pardoen (EHESS), A.-M. Raimond (IGN), S. Robert (EHESS), M.-C. Vouloir (EHESS). 

\section*{Author Contributions}
J.P. took care of the construction of the database and collaborative tools, initiated the project and wrote the paper.
M.G. is responsable for the historical dimension, initiated the project and wrote the paper.
M.B. initiated the project and wrote the paper.

\section*{Competing financial interests}
The author(s) declare no competing financial interests.

\section*{Data Citations}
\label{datacitation1}
1. Perret, J., Gribaudi, M., Barthelemy, M., Abadie, N., Baciocchi, S., Bertelli, C., Bonin, O., Bordin, P., Costes, B., Cristofoli, P., Dumenieu, B., Gravier, J., Hubert, J.-P., Le Ny, P.-A., Mermet, E., Motte, C., Pardoen, M., Raimond, A.-M., Robert, S. \& Vouloir, M.-C., \emph{The 18th century Cassini roads and cities dataset}, http://dx.doi.org/10.7910/DVN/28674, Harvard Dataverse, V5 (2015).


\begin{thebibliography}{1}
\expandafter\ifx\csname url\endcsname\relax
  \def\url#1{\texttt{#1}}\fi
\expandafter\ifx\csname urlprefix\endcsname\relax\def\urlprefix{URL }\fi
\providecommand{\bibinfo}[2]{#2}
\providecommand{\eprint}[2][]{\url{#2}}

\bibitem{cassini}
  \bibinfo{author}{\'Ecole des Hautes Etudes en Sciences Sociales (EHESS) - Laboratoire de d\'emographie et d'histoire sociale},
  \bibinfo{author}{Centre national de la recherche scientifique (CNRS)} \&
  \bibinfo{author}{Biblioth\`eque nationale de France (BnF)},
\newblock \emph{\bibinfo{title}{{Carte de Cassini en couleur (feuilles grav\'ees et aquarell\'ees), issue de l'exemplaire dit de <<Marie-Antoinette>> du XVIIIe si\`ecle}}}
  (\bibinfo{year}{1999}).

\bibitem{geoportail}
  \bibinfo{author}{Institut National de l'Information G\'eographique et Foresti\`ere},
\newblock \emph{\bibinfo{title}{Cassini map on the G\'eoportail}},
\newblock \bibinfo{url}{{http://geoportail.fr/url/7F7dsq}}
  (\bibinfo{year}{2015}).

\bibitem{NYPL}
  \bibinfo{author}{NYPL Labs},
\newblock \emph{\bibinfo{title}{Home page}},
\newblock \bibinfo{url}{{http://www.nypl.org/collections/labs}}
  (\bibinfo{year}{2015}).

\bibitem{Strano2012}
  \bibinfo{author}{Strano, E.},
  \bibinfo{author}{Nicosia, V.},
  \bibinfo{author}{Latora, V.},
  \bibinfo{author}{Porta, S.} \&
  \bibinfo{author}{Barthelemy, M.},
\newblock \bibinfo{title}{{``Elementary processes governing the evolution of road networks''}},
\newblock \emph{\bibinfo{journal}{Scientific reports}},
  \textbf{\bibinfo{volume}{2}}
  (\bibinfo{year}{2012}).

\bibitem{Barthelemy2013}
  \bibinfo{author}{Barthelemy, M.},
  \bibinfo{author}{Bordin, P.},
  \bibinfo{author}{Berestycki, H.} \&
  \bibinfo{author}{Gribaudi, M.},
\newblock \bibinfo{title}{{``Self-organization versus top-down planning in the evolution of a city''}},
\newblock \emph{\bibinfo{journal}{Scientific reports}},
  \textbf{\bibinfo{volume}{3}}
  (\bibinfo{year}{2013}).

\bibitem{Dupouey2007}
  \bibinfo{author}{Dupouey, J.L.},
  \bibinfo{author}{Bachacou, J.},
  \bibinfo{author}{Cosserat, R.},
  \bibinfo{author}{Aberdam, S.},
  \bibinfo{author}{Vallauri, D.},
  \bibinfo{author}{Chappart, G.} \&
  \bibinfo{author}{Corvisier de Vill\`ele, M.A.},
\newblock \bibinfo{title}{{``Vers la r\'ealisation d'une carte g\'eor\'ef\'erenc\'ee des for\^ets anciennes de France''}},
\newblock \emph{\bibinfo{journal}{Le Monde des Cartes}},
  \textbf{\bibinfo{volume}{191}}
  (\bibinfo{year}{2007}).

\bibitem{Vallauri2012}
  \bibinfo{author}{Vallauri D.},
  \bibinfo{author}{Grel A.},
  \bibinfo{author}{Granier E.} \&
  \bibinfo{author}{Dupouey J.L.},
\newblock \emph{\bibinfo{title}{{Les for\^ets de Cassini. Analyse quantitative et comparaison avec les for\^ets actuelles}}},
\newblock Rapport WWF/INRA, Marseille
  (\bibinfo{year}{2012}).

\bibitem{Masucci2013}
  \bibinfo{author}{Masucci, A.P.},
  \bibinfo{author}{Stanilov, K.} \&
  \bibinfo{author}{Batty, M.},
\newblock \bibinfo{title}{{``Limited Urban Growth: London's Street Network Dynamics since the 18th Century''}},
\newblock \emph{\bibinfo{journal}{PLoS ONE}},
  \textbf{\bibinfo{volume}{8}}(8)
  (\bibinfo{year}{2013}).

\bibitem{Wang2015}
  \bibinfo{author}{Wang, C.},
  \bibinfo{author}{Ducruet, C.} \&
  \bibinfo{author}{Wang, W.},
\newblock \bibinfo{title}{{``Evolution, accessibility and dynamics of road networks in China from 1600 BC to 1900 AD''}},
\newblock \emph{\bibinfo{journal}{Journal of Geographical Sciences}},
  \textbf{\bibinfo{volume}{25}}(4),
  \bibinfo{pages}{451--484}
  (\bibinfo{year}{2015}).

\bibitem{Gribaudi2014}
  \bibinfo{author}{Gribaudi, M.},
\newblock \emph{\bibinfo{title}{{Paris ville ouvri\`ere : une histoire occult\'ee (1789-1848)}}},
  \bibinfo{publisher}{La D\'ecouverte}
  (\bibinfo{year}{2014}).

\bibitem{Porta2014}
  \bibinfo{author}{Porta, S.},
  \bibinfo{author}{Romice, O.},
  \bibinfo{author}{Maxwell, J. A.},
  \bibinfo{author}{Russell, P.} \&
  \bibinfo{author}{Baird, D.},
\newblock \bibinfo{title}{{``Alterations in scale: patterns of change in main street networks across time and space''}},
\newblock \emph{\bibinfo{journal}{Urban Studies}},
  \textbf{\bibinfo{volume}{51}}(16),
  \bibinfo{pages}{3383-3400}
  (\bibinfo{year}{2014}).

\bibitem{Walter1999}
  \bibinfo{author}{Walter, V.} \&
  \bibinfo{author}{Fritsch, D.},
\newblock \bibinfo{title}{{``Matching spatial data sets: a statistical approach''}},
\newblock \emph{\bibinfo{journal}{International Journal of Geographical Information Science}},
  \textbf{\bibinfo{volume}{13}}
  (\bibinfo{year}{1999}).

\bibitem{Olteanu2008}
  \bibinfo{author}{Olteanu-Raimond, A.-M.} \&
  \bibinfo{author}{Musti{\`e}re, S.},
\newblock \bibinfo{title}{{``Data Matching--a Matter of Belief''}},
\newblock \emph{\bibinfo{journal}{Headway in Spatial Data Handling, Lecture Notes in Geoinformation and Cartography}},
\bibinfo{pages}{501--519}
  (\bibinfo{year}{2008}).

\bibitem{Maraldi1744}
  \bibinfo{author}{Giovan Domenico Maraldi} \&
  \bibinfo{author}{C\'esar-Fran\c{c}ois Cassini de Thury},
\newblock \emph{\bibinfo{title}{{Nouvelle carte qui comprend les  principaux triangles qui servent de fondement \`a la description g\'eom\'etrique de la France}}},
  Paris, Delisle
  (\bibinfo{year}{1744}).

\bibitem{geohistoricaldatawebsite}
\bibinfo{author}{{GeoHistoricalData}},
\newblock \emph{\bibinfo{title}{Home page}},
\newblock \bibinfo{url}{{https://www.geohistoricaldata.org/}}
  (\bibinfo{year}{2015}).

\bibitem{postgresql}
\bibinfo{author}{{PostgreSQL}},
\newblock \emph{\bibinfo{title}{Home page}},
\newblock \bibinfo{url}{{http://www.postgresql.org}}
  (\bibinfo{year}{2015}).

\bibitem{postgis}
\bibinfo{author}{{PostGIS}},
\newblock \emph{\bibinfo{title}{Home page}},
\newblock \bibinfo{url}{{http://postgis.net}}
  (\bibinfo{year}{2015}).

\bibitem{qgis}
\bibinfo{author}{{Quantum GIS}},
\newblock \emph{\bibinfo{title}{Home page}},
\newblock \bibinfo{url}{{http://qgis.org}}
  (\bibinfo{year}{2015}).

\bibitem{cassiniwebsite}
\bibinfo{author}{{EHESS}},
\newblock \emph{\bibinfo{title}{Cassini Website}},
\newblock \bibinfo{url}{{http://cassini.ehess.fr}}
  (\bibinfo{year}{2015}).

\bibitem{Pelletier2002}
  \bibinfo{author}{Pelletier, M.} \&
  \bibinfo{author}{Carrez, J.-F.},
\newblock \emph{\bibinfo{title}{{Les cartes des Cassini~: la science au service de l'\'Etat et des r\'egions}}},
  \bibinfo{publisher}{Paris, E. du CTHS}
  (\bibinfo{year}{2002}).

\bibitem{Bonin2014}
  \bibinfo{author}{Bonin, O.},
\newblock \bibinfo{title}{{``Analyse de la croissance de r\'eseaux de transport sur le moyen terme \`a partir de sources cartographiques''}},
\newblock \bibinfo{booktitle}{\emph{Croissance et d{\'e}croissance des r{\'e}seaux}},
\newblock \bibinfo{editor}{Beauguitte, L.}, ed.,
\newblock \bibinfo{url}{{https://halshs.archives-ouvertes.fr/halshs-01068589}}
  (\bibinfo{year}{2014}).

\bibitem{postgistopology}
\bibinfo{author}{{PostGIS Topology}},
\newblock \emph{\bibinfo{title}{Topology Manual}},
\newblock \bibinfo{url}{{http://postgis.net/docs/manual-dev/Topology.html}}
  (\bibinfo{year}{2015}).

\bibitem{cassinitopology}
\bibinfo{author}{{Julien Perret}},
\newblock \emph{\bibinfo{title}{{cassini-topology}}},
\newblock \bibinfo{url}{{http://dx.doi.org/10.6084/m9.figshare.1515888}}
  (\bibinfo{year}{2015}).

\bibitem{buildinginspector}
\bibinfo{author}{NYPL Labs},
\newblock \emph{\bibinfo{title}{{Building Inspector}}},
\newblock \bibinfo{url}{{http://buildinginspector.nypl.org}}
  (\bibinfo{year}{2015}).

\bibitem{arpenteurtopographe}
\bibinfo{author}{GeoHistoricalData},
\newblock \emph{\bibinfo{title}{{L'Arpenteur Topographe}}},
\newblock \bibinfo{url}{{http://geohistoricaldata.herokuapp.com}}
  (\bibinfo{year}{2014}).

\bibitem{arpenteurtopographe-code}
\bibinfo{author}{NYPL} \&
\bibinfo{author}{GeoHistoricalData},
\newblock \emph{\bibinfo{title}{{Code for ``L'Arpenteur Topographe''}}},
\newblock \bibinfo{url}{{https://github.com/IGNF/building-inspector}}
  (\bibinfo{year}{2014}).

\end{thebibliography}
\end{document}